 \documentclass[final,5p,times,twocolumn]{elsarticle}
\usepackage{amsmath}
\usepackage{slashed}

\usepackage{amssymb}

\newcommand{\sfrac}[2]{{\textstyle\frac{#1}{#2}}}

\newcommand{\ihalf}{\sfrac{i}{2}}

\newcommand{\alg}[1]{\mathfrak{#1}}

\journal{Physics Letters B}

\begin{document}

\begin{frontmatter}



\title{Scalar one-point functions and matrix product states of AdS/dCFT }

\author{Marius de Leeuw$^a$, Charlotte Kristjansen$^b$, and {Georgios Linardopoulos $^{c,}$$^d$,}}

\address[label1]{School of Mathematics \& Hamilton Mathematics Institute\\
Trinity College Dublin \\
Dublin, Ireland}

\address[label2]{Niels Bohr Institute, Copenhagen University,\\ Blegdamsvej 17, 2100 Copenhagen \O, Denmark}
\address[label3] {Institute of Nuclear and Particle Physics, N.C.S.R., ``Demokritos",\\
153 10 Agia Paraskevi, Greece.}

 \address[label4] {Department of Physics, National and Kapodistrian University of Athens,\\
Zografou Campus, 157 84 Athens, Greece.}



\begin{abstract}
We determine in a closed form all scalar one-point functions  of the defect CFT dual to the D3-D5 probe brane system with
$k$ units of flux which amounts to calculating the overlap between a Bethe eigenstate of the integrable SO(6) spin chain and a certain matrix product state of bond dimension $k$. In particular, we show that the matrix product state is annihilated by all the parity odd charges of the spin chain which has recently been suggested as the criterion for such a state to correspond
to an integrable initial state. Finally, we discuss the properties of the analogous matrix product state for  the
SO(5) symmetric D3-D7 probe brane set-up.
\end{abstract}


\end{frontmatter}



\section{Introduction}
Matrix product states continue to make their appearance in yet more diverse problems in theoretical physics.  In a series of
recent works matrix product states were used to obtain a convenient reformulation of the problem of calculating one-point functions in certain defect versions of ${\cal N}=4$
SYM~\cite{deLeeuw:2015hxa,Buhl-Mortensen:2015gfd,deLeeuw:2016umh,deLeeuw:2016ofj}.
More precisely, the tree-level one-point functions of the defect
CFT could be expressed as the overlap between a
matrix product state and a Bethe eigenstate of an integrable spin chain.
By a matrix product state we understand a state of the form
\begin{align}
\sum_{\vec{i}}   \mathrm{tr} [A_{i_i}\ldots A_{i_L}]\,\, | e_{i_1} \ldots e_{i_L}\rangle,  \label{MPSgeneral}
\end{align}
where the  $e_i$ take values among the basis vectors in an appropriate Hilbert space and  the $A_j$ are
matrices of size $d\times d$ where $d$ is the bond dimension.
Matrix product states can be considered a special class of initial states for quantum quenches in integrable lattice models
and have been studied from this perspective f.inst.~in~\cite{Mestyan:2017xyk,Piroli:2017sei}. In this
connection the object of interest is simply the overlap between the given initial state and the Bethe eigenstates of the integrable system as these
quantities provide input for the study of the relaxation properties of the system after a quench.  A number of
overlap formulas have appeared for the case of the XXZ Heisenberg spin chain. In~\cite{Pozsgay:2009} formulas for the overlap between a Bethe eigenstate and respectively the N\'{e}el state, the  dimer state and the $q$-deformed dimer state were derived.
Later it was shown that these formulas could be expressed in terms of determinants built out of  Bethe
roots~\cite{Brockmann:2014a,Brockmann:2014b} and the result was generalized to the $m$'th raised version of the N\'{e}el state~\cite{Brockmann2014}. The N\'{e}el state, the  dimer state and the $q$-deformed dimer state are
all examples of so-called two-site product states, i.e.\ states of even length which can be built by tensoring identical two-site states. Recently, an expression for the overlap between any such two-site product state and a Bethe eigenstate of the
XXZ spin chain was presented~\cite{Pozsgay:2018ixm}.

The study of one-point functions of non-protected operators in the defect version of ${\cal N}=4$ SYM dual to the D3-D5 probe brane system with $k$ units of background gauge field flux has sparked the derivation of overlap formulas involving matrix product states. An example of a
matrix product state which is of relevance for these considerations is the following state of the SU(2) or Heisenberg XXX spin chain
\begin{equation}
\label{MPS}
 | \text{MPS} \rangle =\text{tr}\prod_{l=1}^L\Big(\,|\!\uparrow\rangle\, \otimes t_1+|\!\downarrow \rangle \,\otimes t_2\Big),
\end{equation}
where $t_1$ and $t_2$ belong to a set of three $k\times k$ matrices $t_i$  which constitute a $k$-dimensional unitary, irreducible representation of $\mathfrak{su}(2)$,
in particular
\begin{equation}\label{eq:su2relations}
 \left[t_i,t_j\right]=i\varepsilon _{ijk}t_k.
\end{equation}
 In~\cite{deLeeuw:2015hxa} a closed expression of determinant type was found for the overlap between
the matrix product state given in eqn.~(\ref{MPS}) for $k=2$ and a Bethe eigenstate of the SU(2) spin chain.  The result could be proved by
relating the matrix product state via cohomology to the N\'{e}el state or one of its
raised versions~\cite{deLeeuw:2015hxa,Buhl-Mortensen:2015gfd}, see also~\cite{Foda:2015nfk,Piroli:2017sei} for alternative proofs. Furthermore, from the $k=2$ result the result for general $k$ could be derived by recursion~\cite{,Buhl-Mortensen:2015gfd}. A first step in the direction of generalizing the results to integrable spin chains for which the Bethe equations
involve nesting was taken in~\cite{deLeeuw:2016umh} where the overlap formula between a Bethe eigenstate of the SU(3) Heisenberg spin chain and the generalization of the matrix product state~(\ref{MPS}) for $k=2$ was found.  This formula will
be reproduced as a special case below where we present an overlap formula involving the Bethe eigenstates of the
integrable SO(6) spin chain, which has two levels of nesting, and a specific type of matrix product state, encoding
the information about all scalar one-point functions of the D3-D5 probe brane set-up for any value of the flux $k$.

 The discovery of the above mentioned series of exact overlap formulas naturally leads to speculations about what precisely
characterises the corresponding initial states or, stated differently, if for integrable lattice models there exists a certain sub-set of initial states which in some sense are integrable. Reference~\cite{Piroli:2017sei} proposes an integrability criterion applicable to
models such as  integrable spin chains which can be solved by the algebraic Bethe ansatz. Assume that the transfer matrix of
the model has been constructed in such a way that the encoded local conserved charges all have a specific parity. Then the proposal states that a given initial state is integrable if it is annihilated by all the odd charges of the model. The proposal
is inspired by the classical work on continuum 2D field theories~\cite{Ghoshal:1993tm} in which a field theory with a boundary is said to be integrable if an infinite subset of the constants of motion  of the original theory are still conserved when a boundary
Hamiltonian is added.  All the earlier mentioned initial states for which a closed overlap formula has been found fulfill the proposed criterion for integrability but the proposal does not come with any information about the form of the overlap formula
or with a strategy for how to obtain it.
The criterion, however, is easy to apply and we shall show that indeed the initial states for which we find below a closed overlap
formula do fulfill it as well.

We begin in section~\ref{One-pt} by briefly describing the defect version of ${\cal N}=4$ SYM dual to the D3-D5 probe
brane system with flux as well as the strategy for computing its tree-level one-point functions using matrix product states.
We shall be brief and refer to~\cite{deLeeuw:2017cop} for details.
Subsequently, we demonstrate
in section~\ref{Int_proof} that the matrix product state capturing the complete set of scalar one-point functions of this defect CFT fulfills the integrability criterion put forward in~\cite{Piroli:2017sei}. In section~\ref{Results} we present a closed expression of
determinant form for all tree-level scalar one-point functions of the defect CFT, valid for any value of the flux parameter $k$.
Section~\ref{D3-D7} contains a discussion of one-point functions and matrix product states for another
defect CFT, generated by
holography from a D3-D7 probe brane system hosting a non-trivial instanton number. Finally, section~\ref{Conclusion} contains
our conclusion.

\section{One Point Functions and Matrix Product States\label{One-pt}}

Holographic dualities relating probe brane systems and defect conformal field theories can be engineered using the Karch-Randall construction~\cite{Karch:2000gx}. The defect versions of ${\cal N}=4$ SYM which we shall consider in this letter are
all obtained in this
way~\cite{DeWolfe:2001pq,Erdmenger:2002ex,Constable:2001ag}.
We will mainly be concerned with the defect CFT which is dual to the D3-D5 probe brane system where
the D5 probe has geometry $AdS_4\times S^2$
and where  $k$ units of gauge field flux is carried by the $S^2$\cite{Constable:1999ac,Constable:2001ag}. This defect
CFT consists of ${\cal N}=4$ SYM with different ranks of the gauge group on the two sides of a co-dimension one defect
placed at, say, $x_3=0$. This difference in rank is achieved by assigning to three of the six scalar fields of ${\cal N}=4$ SYM a
non-vanishing and space-time dependent vacuum expectation value on one side of the defect.
More precisely, for $x_3>0$
\begin{align}\label{Phiclass}
&\phi _i^{\rm cl} = -
 \frac{1}{x_3}\,\begin{pmatrix}
  \left(t_i\right)_{k\times k} & 0_{k\times (N-k)}\\
  0_{(N-k)\times k} & 0_{(N-k)\times (N-k)} \\
 \end{pmatrix},&&i=1,2,3,\\
&\phi ^{\rm cl}_i=0,&& i=4,5,6.
\end{align}
For $x_3<0$
the classical fields (which are matrices of size $(N-k) \times (N-k)$) vanish. The gauge group is hence (broken) SU($N$)
for $x_3>0$ and SU($N-k$) for $x_3<0$. The $t$ matrices form the $k$-dimensional representation of the $\alg{su}(2)$ algebra \eqref{eq:su2relations}.

At one-loop level the scalars of ${\cal N}=4$ SYM constitute a closed sector and the good conformal single trace operators
can be characterized as being Bethe eigenstates of the integrable SO(6) spin chain
 with Lax matrix  \cite{Minahan:2002ve}
\begin{align}\label{eq:LaxSO6}
L(u) = 1 - \frac{i}{u}\,P + \frac{i}{u-2i}\,K,
\end{align}
where $P$ is the permutation operator and $K$ the trace operator.
For this description it is useful
to combine the real scalar fields into complex ones in the following way
\begin{align}
&X = \phi_1 + i \phi_4,
&&Y = \phi_2 + i \phi_5,
&&Z = \phi_3 + i \phi_6, \label{XYZ}\\
&\bar{X} = \phi_1 - i \phi_4,
&&\bar{Y} = \phi_2 - i \phi_5, \label{XYZbar}
&&\bar{Z} = \phi_3 - i \phi_6.
\end{align}
With these definitions one can identify the closed  SU(2) sub-sector as a sector built from two (non-conjugate) complex fields,
say $X$ and $Y$, and the  SU(3) sub-sector as the sector built from three complex fields $X$, $Y$ and $Z$. For these sub-sectors,
the corresponding Lax operators do not contain the trace operator.
Already at tree level some of the conformal operators acquire a non-vanishing one-point function, obtained by replacing the constituent fields by their classical values. The one-point functions take the form characteristic of a
defect CFT~\cite{Cardy:1984bb}
\begin{align}
\langle {\cal O}_{\Delta}\rangle=\frac{C}{x_3^{\Delta}},
\end{align}
where $\Delta$ denotes the conformal dimension of the operator in question which in the present case is equal to the number, $L$, of constituent
fields of the operator.

The constant $C$ can conveniently
be expressed via the overlap between the relevant Bethe eigenstate and a matrix product state
which in the basis of real fields takes the form
\begin{align}\label{MPSreal}
|\mathrm{MPS_k}\rangle  = \sum_{\vec{i}} \mathrm{tr} [t_{i_i}\ldots t_{i_L}] |\phi_{i_1} \ldots \phi_{i_L}\rangle,
\end{align}
where obviously the bond  dimension is equal to $k$, the dimension of the representation for the vevs of the scalar fields.
In more precise terms
\begin{equation}\label{genericCso6}
 C=
 \left(\frac{8\pi ^2}{\lambda }\right)^{\frac{L}{2}}L^{-\frac{1}{2}}
 \,C_k, \hspace{0.5cm} C_k=
 \frac{\left\langle \Psi\,\right.\!\!\left|\vphantom{\Psi}{\rm MPS}_k \right\rangle}{\left\langle \Psi \right.\!\!\left|\Psi  \right\rangle^{\frac{1}{2}}}.
\end{equation}
The pre-factor  ensures the canonical normalization of the two-point functions of ${\cal N}=4$ SYM,
$\lambda$ is the 't Hooft coupling constant and
$|\Psi \rangle$ the Bethe eigenstate.
We refer to the Les Houches lectures~ \cite{deLeeuw:2017cop} for more details of the set-up.

\section{Matrix Product States and Conserved Charges\label{Int_proof}}

As mentioned above, reference~\cite{Piroli:2017sei} proposes that a matrix product state can be characterized as
integrable if it is annihilated by all the charges which are odd under parity. In the following we will show that the matrix
product state~\eqref{MPSreal} fulfills this criterion.
 Let $\sigma$ be the parity operation
\begin{align}
\sigma: v_1\otimes\ldots\otimes v_L  \mapsto v_L \otimes\ldots\otimes v_1.
\end{align}
Furthermore, let $t(\nu)$ be the transfer matrix of the SO(6) spin chain constructed using the Lax operator in
eqn.\ (\ref{eq:LaxSO6}), ensuring that all charges have a
definite  parity. Then the criterion for integrability is equivalent to the following condition~\cite{Piroli:2017sei}
\begin{align}\label{eq:intcriterion}
\sigma t(v) \sigma |\mathrm{MPS}\rangle = t(v) |\mathrm{MPS}\rangle.
\end{align}
Expressing the transfer matrix as  a product of Lax operators \eqref{eq:LaxSO6} and following the idea of \cite{Buhl-Mortensen:2015gfd}, it is easy to show that the action of the transfer matrix on the matrix product state~\eqref{MPSreal}
can be expressed as
\begin{align}
t(v)|\mathrm{MPS}\rangle  = \sum_{\vec{i}} \mathrm{tr} [\tau_{i_i}\ldots \tau_{i_L}] |\phi_{i_1} \ldots \phi_{i_L}\rangle,
\end{align}
where the $\tau$'s are $6k\times 6k$ matrices that take the form
\begin{align}
\tau_A = 1_{6} \otimes t_A \delta_{A=1,2,3} + \frac{i}{v-2i}\, E^A_j \otimes t_j -\frac{i}{v}\, E_A^j \otimes t_j.
\end{align}
It can then be demonstrated (cf.\ \ref{similarity}) that there exists a similarity transformation $U$ such that the $\tau$-matrices behave exactly as the
$t$-matrices under transposition, \textit{i.e.}
\begin{align}
&\{t_1,t_2,t_3\}^T = \{t_1,-t_2,t_3\},
&&\{\tau_1,\tau_2,\tau_3\}^T = U^{-1}\{\tau_1,-\tau_2,\tau_3\} U. \label{transposition}
\end{align}
Now it is straightforward to prove \eqref{eq:intcriterion}
\begin{align}
\sigma t(v) \sigma |\mathrm{MPS}\rangle&=\label{eq:intD3D5}
\sigma t (\nu) \sum_{\vec{i}} \mathrm{tr} [t^T_{i_1}\ldots t^T_{i_L}] |\phi_{i_1} \ldots \phi_{i_L}\rangle   \\ \nonumber
&=\sigma t (\nu)\sum_{\vec{i}} (-1)^{\#t_2}\mathrm{tr} [t_{i_1}\ldots t_{i_L}] |\phi_{i_1} \ldots \phi_{i_L}\rangle  \\ \nonumber
&= \sigma \sum_{\vec{i}} (-1)^{\#t_2}\mathrm{tr} [\tau_{i_1}\ldots \tau_{i_L}] |\phi_{i_1} \ldots \phi_{i_L}\rangle   \\ \nonumber
&=  \sum_{\vec{i}} (-1)^{\#t_2}\mathrm{tr} [\tau^T_{i_1}\ldots \tau^T_{i_L}] |\phi_{i_1} \ldots \phi_{i_L}\rangle  \\ \nonumber
&=  \sum_{\vec{i}} (-1)^{2\#t_2}\mathrm{tr} [\tau_{i_1}\ldots \tau_{i_L}] |\phi_{i_1} \ldots \phi_{i_L}\rangle  \\ \nonumber
&= t(v)|\mathrm{MPS}\rangle,
\end{align}
where we used cyclicity of the trace to cancel the similarity transformation $U$.

\section{All Scalar One-point Functions of the D3-D5 Set-up\label{Results}}

Determining the tree-level one-point functions of the conformal operators of the defect CFT dual to the D3-D5 probe brane system with flux amounts to evaluating the overlap between the eigenstates of the integrable SO(6) spin chain and an appropriate matrix product state. In order to be able to identify the various sub-sectors in a straightforward manner it is convenient
to work in the basis of complex fields as explained in section~\ref{One-pt}. This implies that the basis vectors $e_i$ of the matrix product
state~(\ref{MPSgeneral}) are to be identified with the six complex fields appearing in eqns.~(\ref{XYZ}) and~(\ref{XYZbar}) and the
matrices $A_i$ with the corresponding classical values for these fields.

The Bethe eigenstates which diagonalize the transfer matrix of the SO(6) spin chain
are characterized by three sets of Bethe roots
\begin{align}
\{u_{j}\}_{j=1}^{M}, \hspace{1.0cm}
\{v^{\pm}_{j}\}_{j=1}^{N_\pm}
\end{align}
which must fulfill the nested  Bethe equations
\begin{align}\label{eq:SO6BAE}
1&=
\bigg(\frac{u_i-\ihalf}{u_i+\ihalf}\bigg)^L\prod_{j\neq i}^{M}\frac{u_i-u_j+i}{u_i-u_j-i}\prod_{k=1}^{N_+}\frac{u_i-v^+_{k}-\ihalf}{u_i-v^+_k+\ihalf}\prod_{k=1}^{N_-}\frac{u_i-v^-_{k}-\ihalf}{u_i-v^-_k+\ihalf}, \nonumber\\
1&=  \prod_{l\neq i}^{N_+}\frac{v^+_i - v^+_l + i}{v^+_i-v^+_l-i} \prod_{k=1}^{M}\frac{v^+_i -u_k-\ihalf }{v^+_i -u_k+\ihalf}, \\
1&=  \prod_{l\neq i}^{N_-}\frac{v^-_i - v^-_l + i}{v^-_i-v^-_l-i} \prod_{k=1}^{M}\frac{v^-_i -u_k-\ihalf }{v^-_i -u_k+\ihalf}.\nonumber
\end{align}
where $L$ denotes the length of the chain. Given the Bethe roots the eigenstate can be constructed using for instance the
recipe in~\cite{Basso:2017khq}.
By construction, a Bethe
eigenstate is an eigenstate of all the local charges of the spin chain and it is easy to see that a Bethe state can only have a
non-vanishing overlap with the matrix product state if its eigenvalue under the action of any of the odd charges is
equal to zero. Furthermore, one can argue that the number of momentum carrying roots, $M$, must be even. These
two facts in combination imply
that the momentum carrying  roots must come in pairs $\pm u_i$.\footnote{The proof of this is identical to the proof for
the Heisenberg spin chain which was spelled out in~\cite{Piroli:2017sei} building on results from~\cite{Mukhin:2009}.}
As shown in~\cite{deLeeuw:2016umh} this again implies that the auxiliary
roots must fulfill that $\{\nu_i^{\pm}\}=\{-\nu_i^{\pm}\}$. Thus, if $N_{\pm}$ is even the auxiliary roots $\nu^{\pm}$
must likewise come in pairs with opposite signs. If $N_{+}$ or $N_-$ is odd the corresponding set of roots must contain
a single root of value zero in addition to the set of paired roots~\cite{deLeeuw:2016umh}.

 The experience from the study of one-point functions in the SU(2) sub-sector~\cite{deLeeuw:2015hxa,Buhl-Mortensen:2015gfd} and the SU(3) sub-sector in the special
case $k=2$~\cite{deLeeuw:2016umh} has shown that overlap formulas are likely to be expressible in terms of a few building blocks, namely the
Baxter polynomials and the norm matrix of the model in question. It turns out that by use of exactly this type of building
blocks one can write down for the full scalar SO(6) sub-sector a closed expression for the tree-level
one-point functions which works
for all cases that can be tested with the available computer power.

In order to describe the one-point function we define the standard Baxter Q-functions

\begin{align}
&Q_1(x) = \prod_{i=1}^M (x-u_i),
&&Q_\pm(x) = \prod_{j=1}^{N_\pm} (x-v^\pm_j),
\end{align}
as well as the reduced Baxter Q-functions
\begin{align}
\bar{Q}_\pm(x) = \prod_{j=1;v^\pm_j\neq 0}^{N_\pm} (x-v^\pm_j),
\end{align}
where we omit the zero roots in the product.
Furthermore, we introduce the norm matrix, $G$, of the SO(6) spin chain
\begin{align}
&G\equiv \partial_J \phi_I,
\end{align}
where $I,J = 1,\ldots, M + N_+ + N_-$ and $\phi$ is the norm function obtained by taking the logarithm of the right hand side of the Bethe equations \eqref{eq:SO6BAE}. More precisely, $\phi_I$  is defined so that the equations \eqref{eq:SO6BAE} take the form $1=e^{i\phi_I} $. Due to the pair structure of the Bethe root configurations the determinant of the norm matrix factorizes
in the same way as it was the case for the SU(2) and the SU(3) sub-sectors~\cite{Brockmann:2014a,deLeeuw:2016umh}, i.e.
\begin{align}
&&\det G = \det G_+ \cdot \det G_-.
\end{align}

In terms of these building blocks we can write the one-point functions of the SO(6) sector compactly as
\begin{align}
C^{SO(6)}_k =
\sqrt{
\frac{Q_1(0)Q_1(\frac{i}{2})Q_1(\frac{ik}{2})Q_1(\frac{ik}{2})}{\bar{Q}_+(0)\bar{Q}_+(\frac{i}{2})\bar{Q}_-(0)\bar{Q}_-(\frac{i}{2})}
} \cdot \mathbb{T}_{k-1}(0) \cdot \sqrt{\frac{\det G_+}{\det G_-}},\label{SO6formula}
\end{align}
where
\begin{align}\label{TSO(6)}
\mathbb{T}_n(x) = \sum_{a=-\frac{n}{2}}^{\frac{n}{2}}(x+ia)^L \frac{Q_+(x+ia)Q_-(x+ia)}{Q_1(x+i(a+\frac{1}{2})) Q_1(x+i(a-\frac{1}{2}))}.
\end{align}

The result \eqref{SO6formula} contains as special cases all the one-point functions of the SU(2) sub-sector as well as of the
SU(3) sub-sector. To reduce to one of these smaller sub-sectors one simply has to  set equal to one the
Baxter polynomials referring to roots which are not excited for the given sub-sector. Furthermore, one has to ignore the corresponding entries of the norm matrix.

To be specific one finds in the SU(3) case (discarding the roots $v_j^-$)
\begin{align}
C^{SU(3)}_k = \sqrt{\frac{Q_1(0)Q_1(\frac{i}{2})}{\bar{Q}_+(0)\bar{Q}_+(\frac{i}{2})}} \cdot T_{k-1}(0) \cdot \sqrt{\frac{\det G_+}{\det G_-}},\label{SU3formula}
\end{align}
where
\begin{align}\label{TSU(3)}
T_n(x) = \sum_{a=-\frac{n}{2}}^{\frac{n}{2}}(x+ia)^L \frac{Q_1(x+\frac{i(n+1)}{2}) Q_+(x+ia)}{Q_1(x+i(a+\frac{1}{2})) Q_1(x+i(a-\frac{1}{2}))}.
\end{align}
This result generalizes the result found in~\cite{deLeeuw:2016umh} for the case $k=2$ to the case of any $k$.  For SU(2) the quantity $T_n(x)$ which
appears in the relation for the one-point functions can be identified as the transfer matrix of the SU(2) Heisenberg spin
chain in the $n+1$ dimensional representation. We note, however, that $T_n(x)$ and $\mathbb{T}_n(x)$
appearing in~(\ref{TSU(3)}) and (\ref{TSO(6)}) can not be identified as a transfer matrix of the corresponding spin chains.
Their interpretation constitutes an important open question.

The result~\eqref{SO6formula} can be proven in the case of $(L,2,1,1)$ states. Moreover it
has been checked numerically for the SO(6) states with $(L,M,N_+,N_-) = $ $(4,4,2,2)$, $(6,4,2,2)$, $(5,4,2,1)$, $(7,4,2,1)$, $(7,6,3,2)$ and $(6,6,3,3)$, for $k = 2, \ldots, 6$.
The validity of \eqref{SU3formula} has been checked numerically for SU(3) states with $(L,M,N_+) = $ $(8,4,2)$, $(9,4,1)$, $(10,4,2)$, $(12,6,2)$, $(13,6,3)$, $(12,8,4)$, $(14,8,4)$ and $(16,8,4)$ for all $k = 2, \ldots, 6$.
Checking the most involved case  necessitates a summation over $10^{12}$ terms. The agreement was perfect in all cases.

\section{The D3-D7 Set-up\label{D3-D7}}

There exists another interesting probe brane set-up whose features are very similar to those of the D3-D5 brane set-up studied so far. This is the D3-D7 probe brane system where the probe D7 brane has geometry $AdS_4\times S^4$ and where
the background gauge field has a non-vanishing flux through the $S^4$~\cite{Constable:2001ag}. The field
theory dual is again a defect version of ${\cal N}=4$ SYM where a co-dimension one defect separates two regions
of space-time differing by having unequal vevs for the scalar fields and consequently, in the same way as before, different dimensions of the gauge
group.
Assuming again the defect to be placed at $x_3=0$ the classical fields
take the values~\cite{Constable:2001ag}
\begin{eqnarray} \nonumber
\phi_i ^{\text{cl}}= \frac{G_i \oplus 0_{N-d_G}}{\sqrt{8}\,x_3}, \quad i = 1,\ldots,5, \qquad \phi_6^{\text{cl}} = 0, \hspace{0.5cm} x_3>0,  \label{Solution1}
\end{eqnarray}
whereas they vanish for $x_3<0$. Here,
the $G_i$ are matrices whose commutators generate a $d_G$-dimensional irreducible representation of $\mathfrak{so}(5)$. Such matrices can be constructed starting from the four-dimensional gamma matrices~\cite{Castelino:1997rv}.

Again, one-point functions can be written as an overlap between Bethe eigenstates and a matrix product state
\begin{align}\label{MPS_D7}
|\mathrm{MPS}\rangle  = \sum_{\vec{i}} \mathrm{tr} [G_{i_i}\ldots G_{i_L}] |\phi_{i_1} \ldots \phi_{i_L}\rangle.
\end{align}
Using the symmetry properties of the $G$-matrices it was shown in~\cite{deLeeuw:2016ofj}
that only operators with the following quantum numbers can have non-vanishing one-point functions
\begin{align}
(L,M,N_+,N_-)= (L,M,M/2,M/2),  \hspace{0.5cm} M \mbox{ even}.
\end{align}
While a closed formula for the non-vanishing one-point functions of the D3-D7 set-up has so far evaded discovery, one can show that the matrix product state~(\ref{MPS_D7}) fulfills the integrability criterion~\eqref{eq:intcriterion} put forward in~\cite{Piroli:2017sei}.  The strategy of the proof is the same as for the D3-D5 case, cf.\ section~\ref{Int_proof}.
 The action of the Lax operator on the $G$-matrices will generate some matrices $\Gamma$ that, up to a similarity transformation, have the same transposition properties as the original $G$-matrices, i.e.
\begin{align}
\{G_1,G_2,G_3,G_4,G_5\}^T = \{-G_1,G_2,-G_3,G_4,G_5\}.
\end{align}
The proof of \eqref{eq:intcriterion} for the D3-D7 set-up is then a copy of \eqref{eq:intD3D5}. (For simplicity, we leave out
the precise expression for the similarity transformation.)

Given that the D3-D7 MPS fulfills the integrability criterion put forward in~\cite{Piroli:2017sei} it would be important for the
applicability of the criterion that a closed expression would exist for the overlap with the Bethe eigenstates and we consider
the continued search for such a formula an important endeavour.

\section{Discussion and Conclusion\label{Conclusion}}

The fact that a closed formula of determinant form could be found for the complete set of scalar one-point functions of the
D3-D5 probe brane set-up, reproducing all available data, justifies characterizing this one-point function problem as integrable. An understanding of this apparent integrability in terms of scattering theory is, however, lacking.  Ideally, one would like to
describe the matrix product state as an integrable boundary state corresponding to some reflection matrix.
Reference~\cite{Piroli:2017sei} proposes a criterion for a given initial state to be integrable and devises a way to
obtain the corresponding reflection matrix  in the well-studied case of the XXZ spin chain which does not involve nesting.
Obtaining a reflection matrix
encoding the properties of the matrix product state~\eqref{MPSreal} of the SO(6) spin chain and being compatible with integrability constitutes an important open problem. Having at hand such a reflection matrix should make
it possible to prove the determinant formula for the overlap~\eqref{SO6formula} by generalizing the
strategy of~\cite{Pozsgay:2009,Foda:2015nfk}. One could then also dream of extending the study of higher loop integrability of one-point functions which was initiated for the SU(2) sector in~\cite{Buhl-Mortensen:2016pxs,Buhl-Mortensen:2016jqo,Buhl-Mortensen:2017ind}.

A clue to progress could be a better understanding of the quantities ${\mathbb{ T}}_{k-1}(0)$ and $T_{k-1}(0)$ appearing
in the overlap formulas~\eqref{SO6formula} and~\eqref{SU3formula}. In the case of the SU(2) spin chain the corresponding
quantity could be identified as the transfer matrix of the $k$'th dimensional representation. In the nested case the ${\mathbb{ T}}_{k-1}(0)$ and $T_{k-1}(0)$ might also be related to transfer matrices in higher representations f.inst. via some kind of projection.

Another way of making progress could be to try to address the problem from the string theory side. So far, in the
string theory language only one-point functions of chiral primaries have been calculated~\cite{Nagasaki:2012re,Kristjansen:2012tn,Buhl-Mortensen:2015gfd} and the treatment of non-protected operators constitutes another interesting open problem.

Finally, as mentioned above, we consider the continued scrutiny of the one-point functions of the D3-D7 probe brane set-up~\cite{deLeeuw:2016ofj} an important step in completing the picture of integrability in relation to one-point functions and
matrix product states.

 \vspace{0.3cm}

\section*{Acknowledgments}
We thank Alessandro Sfondrini, Erik Wid\'{e}n, Matthias Wilhelm and especially Konstantin Zarembo for useful discussions.
Furthermore, we thank Jan Ambj\o rn for giving us access to his computer system.
C.K.\ was supported by DFF-FNU through grant number DFF-4002-00037. MdL was supported by SFI and the Royal Society for funding under grant UF160578. The research of G.L.\ at N.C.S.R.\ "Demokritos" is supported by the General Secretariat for Research and Technology of Greece and from the European Regional Development Fund (KRIPIS).
\vspace*{0.1cm}

\appendix
\section{The Similarity Transformation\label{similarity}}
The explicit form of  the similarity transformation $U$ entering eqn.~\eqref{transposition} is
\begin{align}
U =& \begin{pmatrix}
\alpha_1 1 & \alpha_2 t_3 & -\alpha_2 t_2 & 0 \\
\alpha_2 t_3  & -\alpha_1 1 & -\alpha_2 t_1  & 0 \\
\alpha_2 t_2  & -\alpha_2 t_1 & \alpha_1 1 & 0 \\
0 & 0 & 0 & 1_{3k\times 3k}
\end{pmatrix} +   \nonumber \\
&
\beta \begin{pmatrix}
t_1^2  & -\ihalf[t_2^2,t_3] & \ihalf[t_3^2,t_2] & 0 \\
-\ihalf[t_1^2,t_3] & -t_2^2  & \ihalf[t_3^2,t_1] & 0 \\
-\ihalf[t_1^2,t_2] & \ihalf[t_2^2,t_1] & t_3^2 & 0\\
0 & 0 & 0 & 0
\end{pmatrix},
\end{align}
where,
\begin{align}
&\alpha_1=-1 + i\frac{k^2-1}{2k} \left(\frac{1}{u+\frac{1}{2} i (k-3)}-\frac{1}{u-\frac{1}{2} i (k+3)}\right),\nonumber\\
& \alpha_2 =\frac{(u-i) (2 u-5 i)}{(u-2 i) \left(u+\frac{1}{2} i (k-3)\right) \left(u-\frac{1}{2} i (k+3)\right)}, \nonumber \\
& \beta = -\frac{2 (u-i)}{(u-2 i) \left(u+\frac{1}{2} i (k-3)\right) \left(u-\frac{1}{2} i (k+3)\right)}.
\end{align}
\vspace*{0.3cm}



 \bibliographystyle{elsarticle-num}
 \bibliography{letter.bib}

\begin{thebibliography}{10}
\expandafter\ifx\csname url\endcsname\relax
  \def\url#1{\texttt{#1}}\fi
\expandafter\ifx\csname urlprefix\endcsname\relax\def\urlprefix{URL }\fi
\expandafter\ifx\csname href\endcsname\relax
  \def\href#1#2{#2} \def\path#1{#1}\fi

\bibitem{deLeeuw:2015hxa}
M.~de~Leeuw, C.~Kristjansen, K.~Zarembo, {One-point functions in defect CFT and
  integrability}, JHEP 08 (2015) 098.
\newblock \href {http://arxiv.org/abs/1506.06958} {\path{arXiv:1506.06958}},
  \href {http://dx.doi.org/10.1007/JHEP08(2015)098}
  {\path{doi:10.1007/JHEP08(2015)098}}.

\bibitem{Buhl-Mortensen:2015gfd}
I.~Buhl-Mortensen, M.~de~Leeuw, C.~Kristjansen, K.~Zarembo, {One-point
  functions in AdS/dCFT from matrix product states}, JHEP 02 (2016) 052.
\newblock \href {http://arxiv.org/abs/1512.02532} {\path{arXiv:1512.02532}},
  \href {http://dx.doi.org/10.1007/JHEP02(2016)052}
  {\path{doi:10.1007/JHEP02(2016)052}}.

\bibitem{deLeeuw:2016umh}
M.~de~Leeuw, C.~Kristjansen, S.~Mori, {AdS/dCFT one-point functions of the
  SU(3) sector}, Phys.Lett. B763 (2016) 197.
\newblock \href {http://arxiv.org/abs/1607.03123} {\path{arXiv:1607.03123}},
  \href {http://dx.doi.org/10.1016/j.physletb.2016.10.044}
  {\path{doi:10.1016/j.physletb.2016.10.044}}.

\bibitem{deLeeuw:2016ofj}
M.~de~Leeuw, C.~Kristjansen, G.~Linardopoulos, {One-point functions of
  non-protected operators in the SO(5) symmetric D3-D7 dCFT}, J.Phys. A50~(25)
  (2017) 254001.
\newblock \href {http://arxiv.org/abs/1612.06236} {\path{arXiv:1612.06236}},
  \href {http://dx.doi.org/10.1088/1751-8121/aa714b}
  {\path{doi:10.1088/1751-8121/aa714b}}.

\bibitem{Mestyan:2017xyk}
M.~{Mesty\'{a}n}, B.~Bertini, L.~Piroli, P.~Calabrese, {Exact solution for the
  quench dynamics of a nested integrable system}, J.Stat.Mech. 1708~(8) (2017)
  083103.
\newblock \href {http://arxiv.org/abs/1705.00851} {\path{arXiv:1705.00851}},
  \href {http://dx.doi.org/10.1088/1742-5468/aa7df0}
  {\path{doi:10.1088/1742-5468/aa7df0}}.

\bibitem{Piroli:2017sei}
L.~Piroli, B.~Pozsgay, E.~Vernier, {What is an integrable quench?}, Nucl.Phys.
  B925 (2017) 362.
\newblock \href {http://arxiv.org/abs/1709.04796} {\path{arXiv:1709.04796}},
  \href {http://dx.doi.org/10.1016/j.nuclphysb.2017.10.012}
  {\path{doi:10.1016/j.nuclphysb.2017.10.012}}.

\bibitem{Pozsgay:2009}
B.~Pozsgay, {Overlaps between eigenstates of the XXZ spin-1/2 chain and a class
  of simple product states}, J.Stat.Mech. 2014~(6) (2014) P06011.
\newblock \href {http://arxiv.org/abs/1309.4593} {\path{arXiv:1309.4593}}.

\bibitem{Brockmann:2014a}
M.~Brockmann, J.~De~Nardis, B.~Wouters, J.-S. Caux, {A Gaudin-like determinant
  for overlaps of {N\'{e}el} and XXZ Bethe States}, J.Phys. A47 (2014) 145003.
\newblock \href {http://arxiv.org/abs/1401.2877} {\path{arXiv:1401.2877}}.

\bibitem{Brockmann:2014b}
M.~Brockmann, J.~De~Nardis, B.~Wouters, J.-S. Caux, {{N\'{e}el}-XXZ state
  overlaps: odd particle numbers and Lieb-Liniger scaling limit}, J.Phys. A47
  (2014) 345003.
\newblock \href {http://arxiv.org/abs/1403.7469} {\path{arXiv:1403.7469}}.

\bibitem{Brockmann2014}
M.~Brockmann, {Overlaps of $q$-raised {N\'eel} states with XXZ Bethe states and
  their relation to the Lieb-Liniger Bose gas}, J.Stat.Mech. 5 (2014) P05006.
\newblock \href {http://arxiv.org/abs/1402.1471} {\path{arXiv:1402.1471}}.

\bibitem{Pozsgay:2018ixm}
B.~Pozsgay, {Overlaps with arbitrary two-site states in the XXZ spin chain},
  ~\href {http://arxiv.org/abs/1801.03838} {\path{arXiv:1801.03838}}.

\bibitem{Foda:2015nfk}
O.~Foda, K.~Zarembo, {Overlaps of partial {N\'eel} states and Bethe states},
  J.Stat.Mech. 1602~(2) (2016) 023107.
\newblock \href {http://arxiv.org/abs/1512.02533} {\path{arXiv:1512.02533}},
  \href {http://dx.doi.org/10.1088/1742-5468/2016/02/023107}
  {\path{doi:10.1088/1742-5468/2016/02/023107}}.

\bibitem{Ghoshal:1993tm}
S.~Ghoshal, A.~B. Zamolodchikov, {Boundary S-matrix and boundary state in
  two-dimensional integrable quantum field theory}, Int.J.Mod.Phys. A9 (1994)
  3841, [Erratum: Int.J.Mod.Phys. A9 (1994) 4353].
\newblock \href {http://arxiv.org/abs/hep-th/9306002}
  {\path{arXiv:hep-th/9306002}}, \href
  {http://dx.doi.org/10.1142/S0217751X94001552}
  {\path{doi:10.1142/S0217751X94001552}}.

\bibitem{deLeeuw:2017cop}
M.~de~Leeuw, A.~C. Ipsen, C.~Kristjansen, M.~Wilhelm, {Introduction to
  integrability and one-point functions in $\mathcal{N}=4$ SYM and its defect
  cousin}, in: {Les Houches Summer School: Integrability: From Statistical
  Systems to Gauge Theory Les Houches, France, June 6-July 1, 2016}, 2017.
\newblock \href {http://arxiv.org/abs/1708.02525} {\path{arXiv:1708.02525}}.

\bibitem{Karch:2000gx}
A.~Karch, L.~Randall, {Open and closed string interpretation of SUSY CFT's on
  branes with boundaries}, JHEP 06 (2001) 063.
\newblock \href {http://arxiv.org/abs/hep-th/0105132}
  {\path{arXiv:hep-th/0105132}}, \href
  {http://dx.doi.org/10.1088/1126-6708/2001/06/063}
  {\path{doi:10.1088/1126-6708/2001/06/063}}.

\bibitem{DeWolfe:2001pq}
O.~DeWolfe, D.~Z. Freedman, H.~Ooguri, {Holography and defect conformal field
  theories}, Phys.Rev. D66 (2002) 025009.
\newblock \href {http://arxiv.org/abs/hep-th/0111135}
  {\path{arXiv:hep-th/0111135}}, \href
  {http://dx.doi.org/10.1103/PhysRevD.66.025009}
  {\path{doi:10.1103/PhysRevD.66.025009}}.

\bibitem{Erdmenger:2002ex}
J.~Erdmenger, Z.~Guralnik, I.~Kirsch, {Four-dimensional superconformal theories
  with interacting boundaries or defects}, Phys.Rev. D66 (2002) 025020.
\newblock \href {http://arxiv.org/abs/hep-th/0203020}
  {\path{arXiv:hep-th/0203020}}, \href
  {http://dx.doi.org/10.1103/PhysRevD.66.025020}
  {\path{doi:10.1103/PhysRevD.66.025020}}.

\bibitem{Constable:2001ag}
N.~R. Constable, R.~C. Myers, O.~Tafjord, {Non-abelian brane intersections},
  JHEP 06 (2001) 023.
\newblock \href {http://arxiv.org/abs/hep-th/0102080}
  {\path{arXiv:hep-th/0102080}}, \href
  {http://dx.doi.org/10.1088/1126-6708/2001/06/023}
  {\path{doi:10.1088/1126-6708/2001/06/023}}.

\bibitem{Constable:1999ac}
N.~R. Constable, R.~C. Myers, O.~Tafjord, {The noncommutative bion core}, Phys.
  Rev. D61 (2000) 106009.
\newblock \href {http://arxiv.org/abs/hep-th/9911136}
  {\path{arXiv:hep-th/9911136}}, \href
  {http://dx.doi.org/10.1103/PhysRevD.61.106009}
  {\path{doi:10.1103/PhysRevD.61.106009}}.

\bibitem{Minahan:2002ve}
J.~A. Minahan, K.~Zarembo, {The Bethe ansatz for $\mathcal{N}=4$ super
  Yang-Mills}, JHEP 03 (2003) 013.
\newblock \href {http://arxiv.org/abs/hep-th/0212208}
  {\path{arXiv:hep-th/0212208}}, \href
  {http://dx.doi.org/10.1088/1126-6708/2003/03/013}
  {\path{doi:10.1088/1126-6708/2003/03/013}}.

\bibitem{Cardy:1984bb}
J.~L. Cardy, {Conformal invariance and surface critical behavior}, Nucl.Phys.
  B240 (1984) 514.
\newblock \href {http://dx.doi.org/10.1016/0550-3213(84)90241-4}
  {\path{doi:10.1016/0550-3213(84)90241-4}}.

\bibitem{Basso:2017khq}
B.~Basso, F.~Coronado, S.~Komatsu, H.~T. Lam, P.~Vieira, D.-l. Zhong,
  {Asymptotic four point functions}, ~\href {http://arxiv.org/abs/1701.04462}
  {\path{arXiv:1701.04462}}.

\bibitem{Mukhin:2009}
E.~Mukhin, V.~Tarasov, Sangmin, A.~Varchenko, {Bethe algebra of homogeneous XXX
  Heisenberg model has simple spectrum}, Comm.Math.Phys. 288 (2009) 1.
\newblock \href {http://arxiv.org/abs/0706.0688} {\path{arXiv:0706.0688}},
  \href {http://dx.doi.org/10.1016/S0550-3213(98)00291-0}
  {\path{doi:10.1016/S0550-3213(98)00291-0}}.

\bibitem{Castelino:1997rv}
J.~Castelino, S.~Lee, W.~Taylor, {Longitudinal 5-branes as 4-spheres in matrix
  theory}, Nucl.Phys. B526 (1998) 334.
\newblock \href {http://arxiv.org/abs/hep-th/9712105}
  {\path{arXiv:hep-th/9712105}}, \href
  {http://dx.doi.org/10.1016/S0550-3213(98)00291-0}
  {\path{doi:10.1016/S0550-3213(98)00291-0}}.

\bibitem{Buhl-Mortensen:2016pxs}
I.~Buhl-Mortensen, M.~de~Leeuw, A.~C. Ipsen, C.~Kristjansen, M.~Wilhelm,
  {One-loop one-point functions in gauge-gravity dualities with defects}, Phys.
  Rev. Lett. 117~(23) (2016) 231603.
\newblock \href {http://arxiv.org/abs/1606.01886} {\path{arXiv:1606.01886}},
  \href {http://dx.doi.org/10.1103/PhysRevLett.117.231603}
  {\path{doi:10.1103/PhysRevLett.117.231603}}.

\bibitem{Buhl-Mortensen:2016jqo}
I.~Buhl-Mortensen, M.~de~Leeuw, A.~C. Ipsen, C.~Kristjansen, M.~Wilhelm, {A
  quantum check of AdS/dCFT}, JHEP 01 (2017) 098.
\newblock \href {http://arxiv.org/abs/1611.04603} {\path{arXiv:1611.04603}},
  \href {http://dx.doi.org/10.1007/JHEP01(2017)098}
  {\path{doi:10.1007/JHEP01(2017)098}}.

\bibitem{Buhl-Mortensen:2017ind}
I.~Buhl-Mortensen, M.~de~Leeuw, A.~C. Ipsen, C.~Kristjansen, M.~Wilhelm,
  {Asymptotic one-point functions in gauge-string duality with defects},
  Phys.Rev.Lett. 119~(26) (2017) 261604.
\newblock \href {http://arxiv.org/abs/1704.07386} {\path{arXiv:1704.07386}},
  \href {http://dx.doi.org/10.1103/PhysRevLett.119.261604}
  {\path{doi:10.1103/PhysRevLett.119.261604}}.

\bibitem{Nagasaki:2012re}
K.~Nagasaki, S.~Yamaguchi, {Expectation values of chiral primary operators in
  holographic interface CFT}, Phys.Rev. D86 (2012) 086004.
\newblock \href {http://arxiv.org/abs/1205.1674} {\path{arXiv:1205.1674}},
  \href {http://dx.doi.org/10.1103/PhysRevD.86.086004}
  {\path{doi:10.1103/PhysRevD.86.086004}}.

\bibitem{Kristjansen:2012tn}
C.~Kristjansen, G.~W. Semenoff, D.~Young, {Chiral primary one-point functions
  in the D3-D7 defect conformal field theory}, JHEP 01 (2013) 117.
\newblock \href {http://arxiv.org/abs/1210.7015} {\path{arXiv:1210.7015}},
  \href {http://dx.doi.org/10.1007/JHEP01(2013)117}
  {\path{doi:10.1007/JHEP01(2013)117}}.

\end{thebibliography}

\vspace*{0.5cm}


\end{document}